\documentclass[aps,twocolumn,prb,eqsecnum,showpacs]{revtex4}
\usepackage{graphicx}
\begin{document}
\draft
\preprint{27 April 2005}
\title{Low-temperature thermodynamics of one-dimensional alternating-spin
       \\
       Heisenberg ferromagnets}
\author{Shoji Yamamoto and Hiromitsu Hori}
\address{Division of Physics, Hokkaido University,
         Sapporo 060-0810, Japan}
\date{27 April 2005}
\begin{abstract}
Motivated by a novel bimetallic chain compound in which alternating
magnetic centers are ferromagnetically coupled, we investigate
thermodynamic properties of one-dimensional spin-$(S,s)$ Heisenberg
ferromagnets both numerically and analytically.
On the one hand, quantum Monte Carlo calculations illuminate the overall
thermal behavior.
The specific heat may exhibit a double-peaked structure at intermediate
temperatures for $S\agt 3s$ in general.
On the other hand, a modified spin-wave theory precisely describes the
low-temperature properties.
Expanding the specific heat and the magnetic susceptibility, we reveal an
analogy and a contrast between mixed-spin ferromagnets and ferrimagnets.
\end{abstract}
\maketitle

\section{Introduction}

   Much effort has been devoted to designing molecular systems ordering
ferromagnetically. \cite{K95}
One possible approach consists of assembling molecular bricks so as to
obtain a low-dimensional system with a nonzero resultant spin in the
ground state and then coupling the chains or the layers again in a
ferromagnetic fashion.
Numerous heterospin chain compounds have been synthesized along this
line.
Gleizes and Verdaguer \cite{G7373} made an attempt to alternate two types
of metal ion along one crystallographic axis with antiferromagnetic
intrachain interaction and obtained a pioneering example of
quasi-one-dimensional ferrimagnets, of formula
$\mbox{Mn}A\mbox{Cu(dto)}_2\mbox{(H}_2\mbox{O)}_3
 \cdot 4.5\mbox{H}_2\mbox{O}$
($A=\mbox{Ni},\mbox{Cu};\ 
  \mbox{dto}=\mbox{dithiooxalato}=\mbox{S}_2\mbox{C}_2\mbox{O}_2$).
Kahn {\it et al.} \cite{K782} synthesized another series of bimetallic
chain compounds
$A\mbox{Cu(pbaOH)(H}_2\mbox{O)}_3\cdot n\mbox{H}_2\mbox{O}$
($A=\mbox{Fe},\mbox{Co},\mbox{Ni},\mbox{Cu};\ 
  \mbox{pbaOH}=2\mbox{-hydroxy-}1,3\mbox{-propylenebis(oxamato)}
 =\mbox{C}_7\mbox{H}_6\mbox{N}_2\mbox{O}_7$), one of which indeed attained
the three-dimensional ferromagnetic order at low temperatures.
\cite{P7428}
Caneschi {\it et al.} \cite{C1756} took an alternative strategy of
bringing into interaction metal ions and stable organic radicals.
This idea was developed toward polymeric chain compounds. \cite{O8067}
The wide variety of chemical explorations stimulated the physical interest
in mixed-spin chains.
\cite{D413,V5144,K3336,B3921,P8894,F14709,T5355,N9031,M5908,I3271}

   Most of the thus-far synthesized heterospin systems are characterized
as ferrimagnets.
Ferromagnetic intrachain coupling is observed in few cases.
In such circumstances,
$\mbox{MnNi(NO}_2\mbox{)}_4\mbox{(en)}_2$
($\mbox{en}=\mbox{ethylenediamine} =\mbox{C}_2\mbox{H}_8\mbox{N}_2$),
\cite{K1530} proved to be a quasi-one-dimensional mixed-spin ferromagnet
\cite{F2639} and caused renewed interest in mixed-spin chains.
Gillon {\it et al.} \cite{G14433} calculated the spin density distribution
by means of the density functional theory and quantitatively visualized
the ferromagnetic nature of the Mn(II)-Ni(II) interaction.
Fukushima {\it et al.} \cite{F174430} performed high-temperature series
expansion of the thermal quantities and argued the magnetic structure
including single-ion anisotropy and interchain exchange coupling.
Now an increasing number of chemists and physicists are taking interest in
heterospin ferromagnets. \cite{L1150,L3400}

   Alternating-spin chains possess elementary excitations of dual aspect.
In the case of antiferromagnetic coupling, the acoustic excitations reduce
the ground-state magnetization and are thus of ferromagnetic nature, while
the optical excitations enhance the ground-state magnetization and are thus
of antiferromagnetic nature.
In the case of ferromagnetic coupling, on the other hand, both excitations
are of ferromagnetic character.
Therefore, the Schottky-type peak of the specific heat and the minimum of
the susceptibility-temperature product, which are both ferrimagnetic
features, are absent from mixed-spin ferromagnets.
Nevertheless, mixed-spin ferromagnets and ferrimagnets behave similarly
at low temperatures, which is the goal of this paper.
Employing a quantum Monte Carlo method \cite{Y11033} and a modified
spin-wave theory, \cite{Y14008} we investigate thermodynamis of
one-dimensional alternating-spin Heisenberg ferromagnets with particular
emphasis on the intrinsic low-temperature properties.

\section{Modified Spin-Wave Scheme}

   We consider two kinds of spins $S$ and $s$ ($S>s$) alternating on a
ring with ferromagnetic exchange coupling between nearest neighbors, as
described by the Hamiltonian
\begin{equation}
   {\cal H}
  =-J\sum_{n=1}^N
    \bigl(
     \mbox{\boldmath$S$}_{n} \cdot \mbox{\boldmath$s$}_{n}
    +\mbox{\boldmath$s$}_{n} \cdot \mbox{\boldmath$S$}_{n+1}
    \bigr).
   \label{E:H}
\end{equation}
Even in one dimension, the conventional spin-wave theory
\cite{B206,H1098,D1217} gives a fine piece of information on the
ground-state correlation. \cite{Y13610,I14024}
As for the thermal quantities, however, the low-temperature series
expansion within the conventional scheme \cite{T233} only reproduces the
leading term of the specific heat and nothing correct for the magnetic
susceptibility. \cite{T2808}
Then Takahashi \cite{T168} modified the spin-wave formalism, imposing a
constraint on the magnetization, and obtained an excellent description of
the low-temperature thermodynamics of low-dimensional ferromagnets.
We develop the modified scheme for mixed-spin ferromagnets.

\begin{figure*}
\centering
\includegraphics[width=150mm]{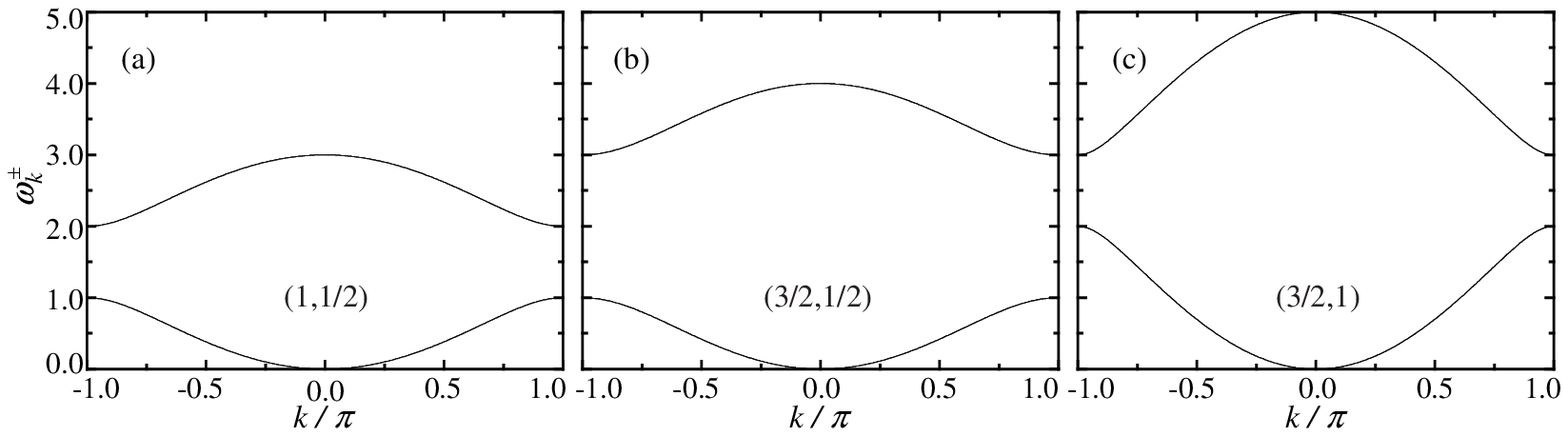}
\caption{Single-magnon excitation spectra as the rigorous dispersion
         relations of the elementary excitations for the spin-$(S,s)$
         ferromagnetic Heisenberg chains.}
\label{F:dsp}
\end{figure*}

   In order to describe the spin deviation in each sublattice, bosonic
operators are introduced as
\begin{equation}
   \left.
   \begin{array}{lll}
      S_n^+=\sqrt{2S-a_n^\dagger a_n}\ a_n,&
      S_n^z=S-a_n^\dagger a_n,\\
      s_n^+=\sqrt{2s-b_n^\dagger b_n}\ b_n,&
      s_n^z=s-b_n^\dagger b_n,
   \end{array}
   \right.
   \label{E:HP}
\end{equation}
where we regard $S$ and $s$ as quantities of the same order.
The bosonic Hamiltonian reads
\begin{equation}
   {\cal H}=E_2+{\cal H}_1+{\cal H}_0+O(S^{-1}),
   \label{E:Hboson}
\end{equation}
where $E_2=-2SsJN$ is the classical ground-state energy and ${\cal H}_i$
is the $O(S^i)$ quantum correction to it.
We consider first diagonalizing ${\cal H}_1$ and then taking ${\cal H}_0$
into calculation perturbationally. \cite{Y211}
Via the transformation
\begin{equation}
   \left.
   \begin{array}{l}
    a_n^\dagger
   ={\displaystyle\frac{1}{\sqrt{N}}\sum_k}{\rm e}^{-{\rm i}k(n-1/4)}
    \bigl(
     \alpha_k^\dagger\cos\theta_k-\beta_k^\dagger\sin\theta_k
    \bigr), \\
    b_n^\dagger
   ={\displaystyle\frac{1}{\sqrt{N}}\sum_k}{\rm e}^{-{\rm i}k(n+1/4)}
    \bigl(
     \alpha_k^\dagger\sin\theta_k+\beta_k^\dagger\cos\theta_k
    \bigr), \\
   \end{array}
   \right.
\end{equation}
with $\tan(2\theta_k)=2\sqrt{Ss}\cos(k/2)/(S-s)$, we obtain
\begin{equation}
   {\cal H}_1=J\sum_k
               \bigl(
                \omega_k^-\alpha_k^\dagger\alpha_k
               +\omega_k^+\beta_k^\dagger\beta_k
               \bigr).
\end{equation}
Here the acoustic ($\omega_k^-$) and optical ($\omega_k^+$) dispersion
relations are given by
\begin{eqnarray}
   &&
   \omega_k^\pm=S+s\pm\sqrt{(S-s)^2+4Ss\cos^2(k/2)}
   \qquad\nonumber \\
   &&\qquad
   \equiv S+s\pm\omega_k,
\end{eqnarray}
and plotted in Fig. \ref{F:dsp}.

   Now we proceed to the modified spin-wave scheme in an attempt to
avoid thermal divergence of the number of bosons.
At finite temperatures, we replace $\alpha_k^\dagger\alpha_k$ and
$\beta_k^\dagger\beta_k$ by
$\bar{n}^\mp_k\equiv\sum_{n^-,n^+}n^\mp P_k(n^-,n^+)$, where
$P_k(n^-,n^+)$ is the probability of $n^-$ acoustic and $n^+$ optical
spin waves appearing in the $k$-momentum state, and minimize the
up-to-$O(S^1)$ free energy
\begin{eqnarray}
   &&
   {\cal F}
  =E_2+J\sum_k\sum_{\sigma=\pm}\omega_k^\sigma\bar{n}_k^\sigma
   \nonumber \\
   &&\qquad
  +k_{\rm B}T\sum_k\sum_{n^-,n^+}P_k(n^-,n^+){\rm ln}P_k(n^-,n^+),
   \qquad
\end{eqnarray}
with respect to $P_k(n^-,n^+)$'s under the condition of zero magnetization
\begin{equation}
   (S+s)N-\sum_k\sum_{\sigma=\pm}\bar{n}_k^\sigma=0,
   \label{E:const}
\end{equation}
together with the trivial constraints $\sum_{n^-,n^+} P_k(n^-,n^+)=1$.
Up to $O(S^1)$, the magnetic susceptibility and the internal energy at
thermal equilibrium are expressed as
\begin{eqnarray}
   &&
   \chi=\frac{(g\mu_{\rm B})^2}{3k_{\rm B}T}
        \sum_k\sum_{\sigma=\pm}\bar{n}_k^\sigma(\bar{n}_k^\sigma+1),
   \qquad \\
   &&
   E=E_2+J\sum_k\sum_{\sigma=\pm}\omega_k^\sigma\bar{n}_k^\sigma,
\end{eqnarray}
with $\bar{n}_k^\pm=[{\rm e}^{(J\omega_k^\pm-\mu)/k_{\rm B}T}-1]^{-1}$,
where the $g$ factors of the spins $S$ and $s$ are both set equal to $g$
and the Lagrange multiplier $\mu$ is determined through the condition
(\ref{E:const}).
The specific heat is calculated by numerically differentiating the
internal energy.
The perturbational correction of $O(S^0)$ reads
\begin{eqnarray}
   &&
   \langle H_0\rangle
  \equiv{\rm Tr}\bigl[{\cal H}_0{\rm e}^{-{\cal H}_1/k_{\rm B}T}\bigr]
       /{\rm Tr}\bigl[{\rm e}^{-{\cal H}_1/k_{\rm B}T}\bigr]
   \nonumber \\
   &&\qquad
  =\frac{JN}{2}
   \biggl[
    \sqrt{\frac{S}{s}}({\mit\Gamma}_1-{\mit\Gamma}_2){\mit\Gamma}_3
   +\sqrt{\frac{s}{S}}({\mit\Gamma}_1+{\mit\Gamma}_2){\mit\Gamma}_3
   \qquad\nonumber \\
   &&\qquad
   -{\mit\Gamma}_1^2+{\mit\Gamma}_2^2-{\mit\Gamma}_3^2
   \biggr],
\end{eqnarray}
with
\begin{eqnarray}
   &&
   {\mit\Gamma}_1=\frac{1}{N}\sum_k(\bar{n}_k^-+\bar{n}_k^+)=S+s,
   \nonumber \\
   &&
   {\mit\Gamma}_2=\frac{1}{N}\sum_k\frac{S-s}{\omega_k}
                  (\bar{n}_k^--\bar{n}_k^+),
   \nonumber \\
   &&
   {\mit\Gamma}_3=\frac{1}{N}\sum_k\frac{2\sqrt{Ss}}{\omega_k}
                  \cos^2\frac{k}{2}(\bar{n}_k^--\bar{n}_k^+),\qquad
\end{eqnarray}
where we keep $\mu$ unchanged.
Indeed $\mu$ may be modified so as to minimize the up-to-$O(S^0)$
free energy, but such a procedure, which is much more complicated, has no
effect on the low-temperature leading behavior. \cite{T233}

\section{Numerical Calculations}

   Quantum Monte Carlo calculations are presented at $N=32$ ($64$ spins).
In the case of $(S,s)=(1,\frac{1}{2})$, we have carried out preliminary
calculations at $N=24,32,40$.
Any quantity divided by $N$ does not vary with $N$ beyond its statistical
error in the temperature range to show.
A few million Monte Carlo steps are spent on low-temperature calculations,
while less than a half million steps on high-temperature calculations.
The numerical precision in the final results is two to three digits.

\begin{figure*}
\centering
\includegraphics[width=150mm]{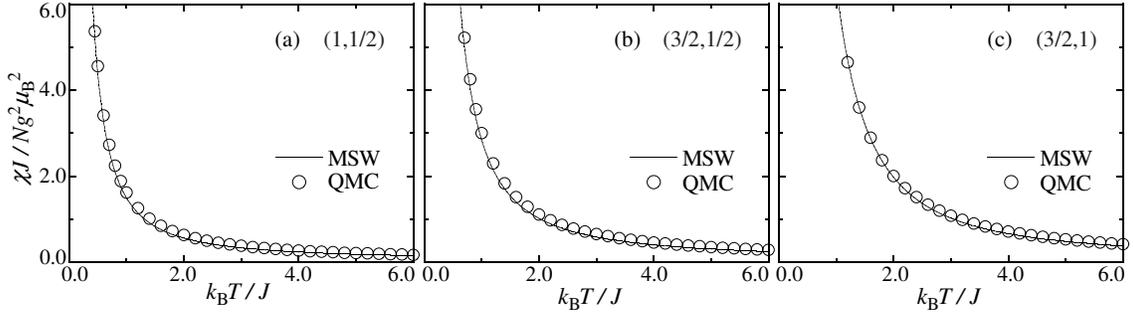}
\caption{Modified-spin-wave (MSW) and quantum Monte Carlo (QMC)
         calculations of the magnetic susceptibility $\chi$ as a function
         of temperature for the spin-$(S,s)$ ferromagnetic Heisenberg
         chains.}
\label{F:chi}
\end{figure*}

   In Fig. \ref{F:chi} we compare modified-spin-wave and quantum Monte
Carlo calculations of the magnetic susceptibility.
They are in excellent agreement over the whole temperature range.
The analytic calculation reproduces the paramagnetic susceptibility
$[S(S+1)+s(s+1)]N(g\mu_{\rm B})^2/3k_{\rm B}T$ at high temperatures and
reveals the $T^{-2}$-diverging behavior at low temperatures, which is
later discussed in more detail.
The susceptibility-temperature product still monotonically decreases with
increasing temperature in contrast with the ferrimagnetic behavior.
\cite{Y1024}

\begin{figure*}
\centering
\includegraphics[width=150mm]{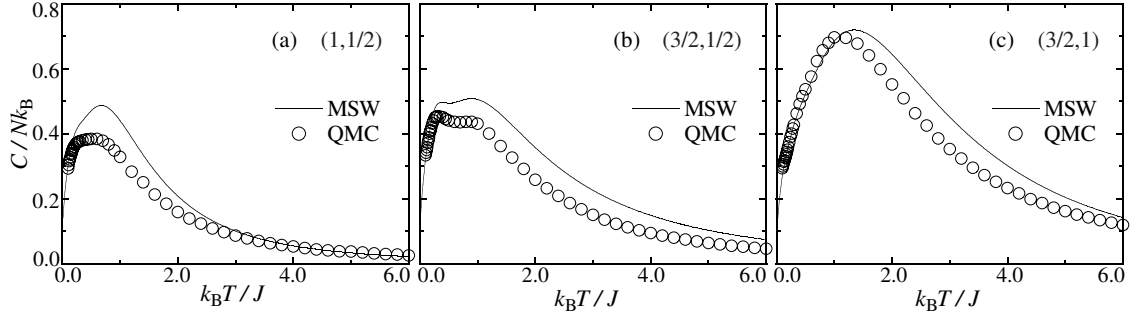}
\caption{Modified-spin-wave (MSW) and quantum Monte Carlo (QMC)
         calculations of the specific heat $C$ as a function of
         temperature for the spin-$(S,s)$ ferromagnetic Heisenberg
         chains.}
\label{F:C}
\end{figure*}

\begin{figure}
\centering
\includegraphics[width=80mm]{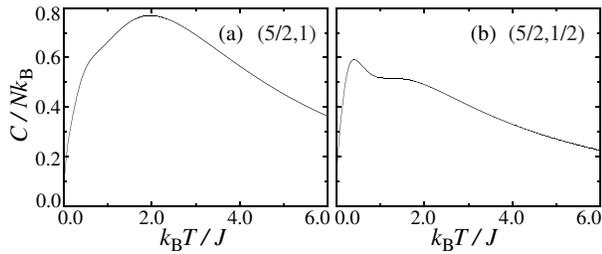}
\caption{Modified-spin-wave (MSW) calculations of the specific heat $C$
         as a function of temperature for the spin-$(S,s)$ ferromagnetic
         Heisenberg chains in the cases of $S<3s$ (a) and $S>3s$ (b).}
\label{F:Cmore}
\end{figure}

   In Fig. \ref{F:C} we compare the modified spin-wave and quantum Monte
Carlo calculations of the specific heat.
The agreement between them is somewhat poorer than that found for the
susceptibility, but the $T^{1/2}$-initial behavior at low temperatures and
the spin-dependent peak structure at intermediate temperatures are well
reproduced by the analytic calculation.
The mid-temperature structure of the specific heat may be regarded as a
function of the acoustic excitation band width $W^-=2sJ$ and the optical
excitation gap ${\mit\Delta}=2SJ$ (see Fig. \ref{F:dsp}).
The heat capacity attributable to the acoustic excitations and that to the
optical excitations may be separable when $W^-\ll{\mit\Delta}$.
Observing further calculations in Fig. \ref{F:Cmore}, we are convinced
that the double-peaked structure may appear for $S\agt 3s$,
including practical cases
$(S,s)
=(\frac{5}{2},\frac{1}{2}),(2,\frac{1}{2}),(\frac{3}{2},\frac{1}{2})$.
Mn(II)Cu(II), Fe(II)Cu(II), and Co(II)Cu(II) chain compounds \cite{K782}
have indeed been synthesized so far, but they all exhibit
antiferromagnetic intrachain interaction.
The double-peaked structure is much more pronounced for ferromagnetic
intrachain interaction. \cite{F174430,N214418}
We expect an increased effort to design ferromagnetic exchange coupling
between alternating metal ions.

\section{Analytical Calculations}

   In order to elucidate the low-temperature thermal behavior, we define
the state density function
\begin{equation}
   w^\pm(x)=\frac{1}{2\pi}\int_{-\pi}^{\pi}
            \delta(x-\omega_k^\pm)\,{\rm d}k.
\end{equation}
Here we are interested in the gapless acoustic branch and expand $w^-(x)$
for small $x$ as
\begin{equation}
   w^-(x)=\frac{1}{\pi}\sqrt{\frac{S+s}{2Ssx}}
          \sum_{n=0}^\infty c_n^- x^n.
   \label{E:w-ex}
\end{equation}
A few leading coefficients are given as
\begin{eqnarray}
   &&
   c_0^-=1,
   \nonumber \\
   &&
   c_1^-=\frac{(S-s)^2+Ss}{4Ss(S+s)},
   \nonumber \\
   &&
   c_2^-=\frac{(3S^2-4Ss+3s^2)(S+s)^2-5S^2s^2}{32S^2s^2(S+s)^2}.\qquad
\end{eqnarray}
Applying Eq. (\ref{E:w-ex}) and neglecting the optical excitations
$\bar{n}_k^+$, Eq. (\ref{E:const}) reads
\begin{eqnarray}
   &&
   v^{1/2}=\frac{1}{\pi\sqrt{2Ss(S+s)}}
           \sum_{n=0}^\infty
           c_n^- t^{n+1/2}
   \nonumber \\
   &&\qquad\times
           {\mit\Gamma}\Bigl(n+\frac{1}{2}\Bigr)
    \biggl[{\mit\Gamma}\Bigl(\frac{1}{2}-n\Bigr)v^n
   \nonumber \\
   &&\qquad
          +\sum_{m=0}^\infty
           \frac{(-1)^m}{m!}\zeta\Bigl(n-m+\frac{1}{2}\Bigr)v^{m+1/2}
    \biggr],\qquad
\end{eqnarray}
where $v=-\mu/k_{\rm B}T$, $t=k_{\rm B}T/J$, and $\zeta(n)$ is Riemann's
zeta function.
Solving this equation iteratively, we obtain the low-temperature expansion
of the Lagrange multiplier as
\begin{eqnarray}
   &&
   v^{1/2}
  =\frac{c_0^-{\mit\Gamma}(1/2)}{\pi\sqrt{2Ss(S+s)}}
   {\mit\Gamma}\Bigl(\frac{1}{2}\Bigr)t^{1/2}
  +\biggl[\frac{c_0^-{\mit\Gamma}(1/2)}{\pi\sqrt{2Ss(S+s)}}\biggr]^2
   \nonumber \\
   &&\qquad\times
   {\mit\Gamma}\Bigl(\frac{1}{2}\Bigr)\zeta\Bigl(\frac{1}{2}\Bigr)t
  +\biggl[\frac{c_0^-{\mit\Gamma}(1/2)}{\pi\sqrt{2Ss(S+s)}}\biggr]^3
   \nonumber \\
   &&\qquad\times
   {\mit\Gamma}\Bigl(\frac{1}{2}\Bigr)
   \biggl[\zeta\Bigl(\frac{1}{2}\Bigr)\biggr]^2 t^{3/2}
  +O(t^2).
\end{eqnarray}
Since the magnetic susceptibility and the internal energy read
\begin{eqnarray}
   &&
   \frac{\chi k_{\rm B}T}{N(g\mu_{\rm B})^2}
  =\frac{1}{3\pi}\sqrt{\frac{S+s}{2Ss}}
   \sum_{n=0}^\infty
   c_n^- t^{n+1/2}
   \nonumber \\
   &&\qquad\times
   {\mit\Gamma}\Bigl(n+\frac{1}{2}\Bigr)
   \biggl[{\mit\Gamma}\Bigl(\frac{3}{2}-n\Bigr)v^{n-3/2}
   \nonumber \\
   &&\qquad
   +\sum_{m=0}^\infty
    \frac{(-1)^m}{m!}\zeta\Bigl(n-m-\frac{1}{2}\Bigr)v^m
   \biggr],\qquad
   \\
   &&
   \frac{E-E_2}{NJ}
  =\frac{1}{\pi}\sqrt{\frac{S+s}{2Ss}}
   \sum_{n=0}^\infty
   c_n^- t^{n+3/2}
   \nonumber \\
   &&\qquad\times
   {\mit\Gamma}\Bigl(n+\frac{3}{2}\Bigr)
   \biggl[{\mit\Gamma}\Bigl(-\frac{1}{2}-n\Bigr)v^{n+1/2}
   \nonumber \\
   &&\qquad
   +\sum_{m=0}^\infty
    \frac{(-1)^m}{m!}\zeta\Bigl(n-m+\frac{3}{2}\Bigr)v^m
   \biggr],\qquad
\end{eqnarray}
the susceptibility and the specific heat are expanded as
\begin{eqnarray}
   &&
   \frac{\chi J}{N(g\mu_{\rm B})^2}
  =\frac{1}{Ss}
   \Biggl\{
    \frac{\tilde{t}^{-2}}{3}
   -\frac{\zeta(1/2)}{\sqrt{2\pi}}\tilde{t}^{-3/2}
   \nonumber \\
   &&\qquad
   +\biggl[\frac{\zeta(1/2)}{\sqrt{2\pi}}\biggr]^2\tilde{t}^{-1}
   \Biggr\}
  +O\bigl(\tilde{t}^{-1/2}\bigr),
   \label{E:MSW-chi}
   \\
   &&
   \frac{C}{Nk_{\rm B}}
  =(S+s)
   \Biggl\{
    \frac{3\zeta(3/2)}{4\sqrt{2\pi}}\tilde{t}^{1/2}
   -\tilde{t}
   \nonumber \\
   &&\qquad
   +\frac{15[(S^2-Ss+s^2)\zeta(5/2)-4\zeta(1/2)]}{32\sqrt{2\pi}}
    \tilde{t}^{3/2}
   \Biggr\}
   \qquad\nonumber \\
   &&\qquad
  +O\bigl(\tilde{t}^2\bigr),
   \label{E:MSW-C}
\end{eqnarray}
where $\tilde{t}=t/Ss(S+s)=k_{\rm B}T/JSs(S+s)$.
The $O(S^0)$ interactions affect the fourth and higher terms
and therefore, whether through the Holstein-Primakoff transformation
\cite{H1098} or through the Dyson-Maleev transformation,
\cite{D1217,M776} we reach the same results (\ref{E:MSW-chi}) and
(\ref{E:MSW-C}).

   Numerically solving the thermodynamic Bethe-ansatz integral equations
for the spin-$\frac{1}{2}$ ferromagnetic Heisenberg chain, Takahashi and
Yamada \cite{T2808} obtained
\begin{eqnarray}
   &&
   \frac{\chi J}{L(g\mu_{\rm B})^2}
  =0.04167t^{-2}+0.145t^{-3/2}+0.17t^{-1}
   \nonumber \\
   &&\qquad
  +O(t^{-1/2}),
   \label{E:BA-chi}
   \\
   &&
   \frac{C}{Lk_{\rm B}}
  =0.7815t^{1/2}-2.00t+3.5t^{3/2}+O(t^2),\qquad
   \label{E:BA-C}
\end{eqnarray}
where $L$ is the number of spins.
When we set $S$ and $s$ both equal to $\frac{1}{2}$, the expressions
(\ref{E:MSW-chi}) and (\ref{E:BA-chi}) coincide in their leading three
terms, while Eqs. (\ref{E:MSW-C}) and (\ref{E:BA-C}) in their leading two
terms.
The modified spin-wave calculations are thus reliable and give rigorous
information on the low-temperature properties.
In the case of arbitrary $S$ and $s$, the leading three terms of Eq.
(\ref{E:MSW-chi}) and the leading two terms of Eq. (\ref{E:MSW-C})
coincide with those of the spin-$[Ss(S+s)/2]^{1/3}$ uniform ferromagnetic
chain except for a common factor.
Considering practical combinations of $S$ and $s$, we may estimate that
$[Ss(S+s)/2]^{1/3}=[1-(S-s)^2/(S+s)^2]^{1/3}(S+s)/2\simeq(S+s)/2$.
Thus, ferromagnetically coupled alternating spins $S$ and $s$ look like
a ferromagnetic assembly of virtual spins $(S+s)/2$ at low temperatures.

   It is also interesting to compare the expressions (\ref{E:MSW-chi}) and
(\ref{E:MSW-C}) with those of ferrimagnetic chains. \cite{Y14008}
It turns out that the spin-$(S,s)$ ferrimagnetic low-temperature
expansions are obtained by replacing $J$ and $s$ by $-J$ and $-s$,
respectively, in Eqs. (\ref{E:MSW-chi}) and (\ref{E:MSW-C}).
In other words, antiferromagnetically coupled alternating spins $S$ and
$s$ look like a ferromagnetic assembly of virtual spins
$[Ss(S-s)/2]^{1/3}=[(S+s)^2/(S-s)^2-1]^{1/3}(S-s)/2$ at low temperatures.
The quantity $[Ss(S-s)/2]^{1/3}$ is less intuitive than the
corresponding $[Ss(S+s)/2]^{1/3}$ in the case of ferromagnetic coupling,
but it results in a realistic spin quantum number $S-s$ when $S$ is equal
to $2s$.
Equation (\ref{E:MSW-chi}) is nothing but the susceptibility of the
spin-$(S,s)$ ferrimagnetic chain if we set $\tilde{t}$ for
$k_{\rm B}T/JSs(S-s)$ instead of $k_{\rm B}T/JSs(S+s)$.
All in all, the low-temperature physics is scaled by $S+s$ in
ferromagnetic chains, whereas by $S-s$ in ferrimagnetic chains.

\section{Concluding Remarks}

   Thermodynamic properties of alternating-spin Heisenberg ferromagnetic
chains have been investigated in comparison with heterospin ferrimagnetic
and homospin ferromagnetic chains.
The magnetic susceptibility is a monotonically decreasing function of
temperature regardless of $(S,s)$ and is qualitatively the same as those
of uniform ferromagnetic chains.
The specific heat qualitatively varies with $(S,s)$ and exhibits a rich
structure at intermediate temperatures.
It may be double-peaked for $S\agt 3s$ in general.

   The low-temperature behavior has been revealed analytically.
The thermal quantities are still expanded in powers of $T^{1/2}$ and
exhibit ferromagnetic features.
The conventional spin-wave theory misunderstands the low-temperature
behavior as series of $T$.
The missing terms are reproduced through the modified procedure.
Ferromagnetic and ferrimagnetic mixed-spin chains are qualitatively alike
at low temperatures.
The spin-$(S,s)$ ferromagnetic chain looks like a ferromagnetic assembly 
of virtual spins
$[Ss(S+s)/2]^{1/3}=[1-(S-s)^2/(S+s)^2]^{1/3}(S+s)/2\simeq(S+s)/2$, while
the spin-$(S,s)$ ferrimagnetic chain behaves like that of virtual spins
$[Ss(S-s)/2]^{1/3}=[(S+s)^2/(S-s)^2-1]^{1/3}(S-s)/2$.
The present findings are really complementary to the sophisticated
high-temperature series-expansion calculations. \cite{F174430,R1390}

   The existent bimetallic chain ferromagnet
$\mbox{MnNi(NO}_2\mbox{)}_4\mbox{(en)}_2$ possesses a rather weak
intrachain exchange coupling ($J/k_{\rm B}\simeq 2\,\mbox{K}$), in which
the low-temperature thermodynamics revealed here is hard to verify.
Nevertheless such a pioneering material must highly motivate further
explorations in both chemical and physical fields, as was the case with
uniform ferromagnetic chains.
\cite{T233,T2808,T168,L463,S2131,T746}
Besides bimetallic chain compounds, several authors \cite{H7921} made a
novel attempt to design low-dimensional heterospin systems utilizing
organic triradicals.
Mixed-spin chains contain further interesting topics such as dynamic
structure factors of dual aspect \cite{Y3711} and nuclear spin relaxation
through the exchange-scattering-enhanced three-magnon process.
\cite{H1453}
We hope our study will stimulate further experimental investigations into
mixed-spin chain compounds.

\acknowledgments

   The authors are grateful to N. Fukushima for valuable comments.
This work was supported by the Ministry of Education, Culture, Sports,
Science and Technology of Japan, and the Iketani Science and Technology
Foundation.

\end{document}